\documentclass[12pt]{iopart}
\usepackage{graphicx}
\usepackage{iopams}
 
\begin{document}

\title{Chaplygin gas in decelerating DGP gravity and the age of the oldest star.}

\author{Matts Roos}

\address{Department of Physical Sciences and Department of
          Astronomy\\ FIN-00014 University of Helsinki, Finland}
\ead{matts.roos@helsinki.fi}
\begin{abstract}
Accelerating Chaplygin gas combined with the decelerating braneworld
Dvali-Gabadadze-Porrati (DGP) model can produce an overall
accelerated expansion of the order of magnitude seen. Both models
have similar asymptotic properties at early and late cosmic times,
and are characterized by a length scale. Taking the length scales to
be proportional one obtains a combined model with three free
parameters, one more than the $\Lambda$CDM model, which fits
supernovae data equally well. We further constrain it by the CMB
shift parameter, and by requiring that the model yields a longer age
of the Universe than that of the oldest star HE 1523-0901, $t_* =
13.4\pm 0.8\,(stat)\pm 1.8\,(syst)$. In contrast to generalized DGP
and Chaplygin gas models, this is a genuine alternative to the
cosmological constant model because it does not reduce to it in any
limit of the parameter space.
\end{abstract}
\maketitle

\section{Introduction}
The demonstration by SNeIa observations that the Universe is
undergoing an accelerated expansion has stimulated a vigorous search
for models to explain this unexpected fact. Since the dynamics of
the Universe is conventionally described by the
Friedmann--Lema\^{\i}tre equations which follow from the Einstein
equation in four dimensions, all modifications ultimately affect the
Einstein equation. The Einstein tensor $G_{\mu\nu}$ encodes the
geometry of the Universe, the stress-energy tensor $T_{\mu\nu}$
encodes the energy density. Thus modifications to $G_{\mu\nu}$ imply
some alternative geometry, modifications to $T_{\mu\nu}$ involve new
forms of energy densities, that have not been observed, and which
therefore are called dark energy.

The traditional solution to this is the cosmological constant
$\Lambda$ which can be interpreted either as a modification of the
geometry or as a vacuum energy term in $T_{\mu\nu}$. This fits
observational data well, in fact no competing model fits data
better. But the problems with $\Lambda$ are well known: its
infinitesimally small value cannot be calculated theoretically in
Quantum Field Theory, and if it can be calculated in string
theories, these can be chosen in a nearly infinite number of ways,
none of which have made any testable predictions.

The search for alternatives to the cosmological constant model
therefore goes on. No modified gravity models nor dark energy models
have been strikingly successful in explaining the cosmic
acceleration, except (at best) by introducing increased complexity
or adding further free parameters. In this situation we think it may
be worthwhile to try to introduce two modifications at the same time
if it can be done economically.

A simple and well-studied model of modified gravity is the
Dvali--Gabadadze--Porrati (DGP) braneworld model (Dvali \&
al.\,\cite{Dvali}, Deffayet \& al.\,\cite{Deffayet}) in which our
four-dimensional world is an FRW brane embedded in a
five-dimensional Minkowski bulk. The model is characterized by a
cross-over scale  $r_c$ such that gravity is a four-dimensional
theory at scales $a\ll r_cH_0$ where matter behaves as pressureless
dust. In the self-accelerating DGP branch, gravity "leaks out" into
the bulk when $a\approx r_cH_0$, and at scales $a \gg r_cH_0$ the
model approaches the behavior of a cosmological constant. To explain
the accelerated expansion which is of recent date ($z\approx 0.5$ or
$a\approx 2/3$), $r_cH_0$ must be of the order of unity. In the
self-decelerating DGP branch, gravity "leaks in" from the bulk at
scales $a\approx r_cH_0$, in conflict with the observed dark energy
acceleration. Note that the self-accelerating branch has a ghost,
whereas the self-decelerating branch is ghost-free.

A simple and well-studied model of dark energy introduces into
$T_{\mu\nu}$ the density $\rho_{\varphi}$ and pressure $p_{\varphi}$
of an ideal fluid called Chaplygin gas (Kamenshchik \&
al.\,\cite{Kamenshchik}, Bili$\acute{\rm c}$ \& al.\,\cite{Bilic})
following Chaplygin's historical work in aerodynamics
\cite{Chaplygin}. Like the previous model, it is also characterized
by a length scale below which the gas behaves as pressureless dust,
at late times approaching the behavior of a cosmological constant.

Both the self-accelerating DGP model and the standard Chaplygin gas
model have problems fitting present observational data. This has
motivated generalizations to higher-dimensional braneworld models
which have at least one parameter more than $\Lambda$CDM, yet they
fit data best in the limit where they reduce to $\Lambda$CDM.

Here we combine the 2-parametric self-decelerating DGP model with
the likewise 2-parametric standard Chaplygin gas model because of
the similarities in their asymptotic properties, taking the length
scales in the models to be proportional. The proportionality
constant subsequently disappears because of a normalizing condition
at $z=0$. Thus the model has only one parameter more than the
standard $\Lambda$CDM model. It is a genuine alternative to the
cosmological constant model because it does not reduce to it in any
limit of the parameter space.

This paper is organized as follows. In Section 2 we discuss the
length scales and parameters in the DGP and Chaplygin gas models as
was first done in Roos\,\cite{Roos1} and developed further in the
references \cite{Roos2,Roos3}. In Section 3 we discuss the basic DGP
model in flat space, the standard Chaplygin gas model, and their
amalgamation. In Section 4 we discuss data, analyses, and fits. In
Section 5 we discuss a constraint on the age of the Universe by
comparing model predictions with the age of the oldest star. In
Section 6 we turn to the dynamical quantities $w_{\rm eff}$ and $q$,
and study their redshift dependences. In Section 7 we discuss the
results and conclude.

\section{Length scales}

On the four-dimensional brane in the DGP model, the action
of gravity is proportional to $M^2_{\rm Pl}$ whereas in the bulk it is
proportional to the corresponding quantity in 5 dimensions, $M^3_5$.
The cross-over length is defined as in Ref.\,\cite{Deffayet},
\begin{equation}
r_c=M^2_{\rm Pl}/2M^3_5\ .\label{rc}
\end{equation}
It is customary to associate a density parameter to this,
\begin{equation}
\Omega_{r_c}=(2r_c H_0)^{-2},\label{Omrc}
\end{equation}
such that $r_c H_0$ is a length scale (similar to $a$).

The Friedmann--Lema\^{\i}tre equation in the DGP model may be written \cite{Deffayet}
\begin{equation}
H^2-\frac k{a^2}-\epsilon\frac 1 r_c\sqrt{H^2-\frac
k{a^2}}=\kappa\rho~,\label{Friedm}
\end{equation}
where $a=(1+z)^{-1},\ \kappa=8\pi G/3$, and $\rho$ is the total
cosmic fluid energy density with components $\rho_m$ for baryonic
and dark matter, and $\rho_{\varphi}$ for whatever additional dark
energy may be present, in our case the Chaplygin gas. Clearly the
standard FLRW cosmology is recovered in the limit
$r_c\rightarrow\infty$. In the following we shall only consider
$k=0$ flat-space geometry. The \emph{self-accelerating branch}
corresponds to $\epsilon=+1$; we shall in the following consider
only the \emph{self-decelerating branch} with $\epsilon=-1$. Since
ordinary matter does not interact with Chaplygin gas, one has
separate continuity equations for the energy densities $\rho_m$ and
$\rho_{\varphi}$, respectively. In DGP geometry the continuity
equations for ideal fluids have the same form as in FLRW geometry
\cite{Deffayet},
\begin{equation}
\dot\rho+3H(\rho+p)=0\ .\label{4}
\end{equation}
Pressureless dust with $p=0$ then evolves as $\rho_m(a)\propto
a^{-3}$. The free parameters in the DGP model are $\Omega_{r_c}$ and
$\Omega_m=\kappa\rho_m/H_0^2$. Note that there is no curvature term $\Omega_k$ since we
have assumed flatness by setting $k=0$ in equation (\ref{Friedm}).

The Chaplygin gas has the barotropic equation of state
$p_{\varphi}=-A/\rho_{\varphi}$ \cite{Kamenshchik,Bilic}, where $A$
is a constant with the dimensions of energy density squared. The
continuity equation (\ref{4}) is then $
\dot\rho_{\varphi}+3H\left(\rho_{\varphi}-A/\rho_{\varphi}\right)=0$,
which integrates to
\begin{equation}
\rho_{\varphi}(a)=\sqrt{A+B/a^6}~,\label{rhoch}
\end{equation}
and where $B$ is an integration constant. Thus this model has two free
parameters. Obviously the limiting behavior of the energy density is
\begin{equation}
\rho_{\varphi}(a)\propto\frac{\sqrt{B}}{a^{3}}~~ {\rm for}~~ a~\ll
\left(\frac B A\right)^{1/6},~\rho_{\varphi}(a)\propto \sqrt{A}~~
{\rm for}~~ a\gg \left(\frac B A\right)^{1/6}.
\end{equation}
In models combining DGP gravity and Chaplygin gas dark energy
\cite{Bouhmadi, Roos2, Roos3} there are thus four free parameters,
$\Omega_{r_c},~\Omega_m,~A$, and $B$, one of which shall be
eliminated in the next Section. We now choose the two length scales,
$r_cH_0$ and $(B/A)^{1/6}$, to be proportional by a factor $x$, so
that
\begin{equation}
\left(\frac B A\right)^{1/6}=xr_c H_0=\frac{x}{2\sqrt{\Omega_{r_c}}}\ .\label{BA}
\end{equation}

It is convenient to replace the parameters $A$ and $B$ in
Eq.\,(\ref{rhoch}) by two new parameters,
$\Omega_A=H_0^{-2}\kappa\sqrt{A}$ and
$x=2\sqrt{\Omega_{r_c}}(B/A)^{1/6}$. The dark energy density can
then be written
\begin{equation}
\rho_{\varphi}(a)=H_0^2\kappa^{-1}\Omega_A \sqrt{1+x^6(4\Omega_{r_c}a^2)~^{-3}}~.\label{rhophi}
\end{equation}

\section{The combined model}

Let us now return to Equation\,(\ref{Friedm}) and solve it for
the expansion history $H(a)$. Substituting $\Omega_{r_c}$ from
Eq.\,(\ref{Omrc})~, $\rho_{\varphi}(a)$ from Eq.\,(\ref{rhophi}), and
$\Omega_m=\Omega_m^0 a^{-3}$, it becomes
\begin{equation}
\frac {H(a)}{H_0}=-\sqrt{\Omega_{r_c}}+\left[\Omega_{r_c}+\Omega_m^0 a^{-3}
+\Omega_A\sqrt{1+x^6(4\Omega_{r_c}a^2)~^{-3}}\right]^{1/2}.\label{Ha}
\end{equation}
Note that $\Omega_{r_c}$ and $\Omega_A$ do not evolve with $a$, just
like $\Omega_{\Lambda}$ in the the $\Lambda$CDM model. In the limit
of small $a$  this equation reduces to two terms which evolve as
$a^{-3/2}$, somewhat similarly to dust with density parameter
$\sqrt{\Omega_m^0 +\Omega_Ax^3(4\Omega_{r_c})^{-3/2}}$. In the limit
of large $a$, Eq.\,(\ref{Ha}) describes a de Sitter acceleration
with a cosmological constant $\Omega_{\Lambda}=
-\sqrt{\Omega_{r_c}}+\sqrt{\Omega_{r_c}+\Omega_A}$.

A closer inspection of Eq.~(\ref{Ha}) reveals that it is not
properly normalized at $a=1$  to ${H(1)}/{H_0}=1$, because the
right-hand-side takes different values at different points in the
space of the parameters $\Omega_m^0,~\Omega_{r_c},\Omega_A$, and
$x$. This gives us a condition: at $a=1$ we require that $H(1)=xH_0$
so that Eq.~(\ref{Ha}) takes the form of a 6:th order algebraic
equation in the variable $x$
\begin{equation}
x=-\sqrt{\Omega_{r_c}}+\left[\Omega_{r_c}+\Omega_m^0
+\Omega_A\sqrt{1+x^6(4\Omega_{r_c})~^{-3}}\right]^{1/2}.\label{Hx}
\end{equation}

This condition shows that $x$ is a function
$x=f(\Omega_m^0,~\Omega_{r_c},\Omega_A)$. Finding real, positive
roots $x$ and substituting them into Eq.~(\ref{Ha}) would normalize
the equation properly. The only problem is that the function cannot
be expressed in closed form, so one has to resort to numerical
iterations. The average value of $x$ is found to be $x\approx
0.956$; it varies over the interesting part of the parameter space,
but only by $\approx 0.002$.

\section{Data, analysis, and fits}
The data we use to test this model are the same 192 SNeIa data as in the compilation used by Davis \& al.\,\cite{Davis} which is a combination of the "passed" set in Table 9 of Wood-Vasey \& al.\,\cite{Wood} and the "Gold" set in Table 6 of Riess \& al.\,\cite{Riess}.

We are sceptical about using CMB and BAO power spectra, because they
have been derived in FRW geometry, not in five-dimensional brane geometry.
The SNeIa data are, however, robust in our analysis, since the distance moduli
are derived from light curve shapes and fluxes, that do not depend on the choice of
cosmological models.

The Davis \& al. compilation \cite{Davis} lists magnitudes $\mu_i$,
magnitude errors $\Delta\mu_i$ for SNeIa at redshifts $z_i,\
i=1,192$. We compute model magnitudes
\begin{equation}
\mu(z_i,\Omega_m,\Omega_{r_c},\Omega_A)= 5~ \rm{Log}\,[d_L(z_i,\Omega_m,\Omega_{r_c},\Omega_A)]+25\ ,
\end{equation}
where the luminosity distance in Mpc at redshift $z_i$ is
\begin{equation}
d_L(z_i,\Omega_m,\Omega_{r_c},\Omega_A)=\frac{(1+z_i)c}{H_0}\int_0^{z_i}\frac{dz}{xH(z)},
\end{equation}
where $H(z)$ is given by Eq.~(\ref{Ha}), and $H_0=72$\,km/(s Mpc).

We then search in the parameter space for a minimum of the $\chi^2$ sum
\begin{equation}
\chi^2=\sum_{i=1}^{192} \left(\frac{\mu_i-\mu(z_i,\Omega_m,\Omega_{r_c},\Omega_A)}{\Delta\mu_i}\right)^2 .\label{chi}
\end{equation}

As is well known in the $\Lambda$CDM model, the supernova data alone
do not determine neither $\Omega_m$ nor $\Omega_{\Lambda}$ well
because they are strongly correlated. What the supernova data
determine well is $\Omega_{\Lambda}-\Omega_m$, but they have
essentially no information on $\Omega_{\Lambda}+\Omega_m$.

The situation here is similar: all the three parameters are strongly
degenerate, and what is determined best is $\Omega_A-\Omega_{r_c}$.
Since no errors can be obtained because of the correlations, some
further constraint is needed to break the degeneracy. One way to do
that is to include as a weak CMB prior on $\Omega_m^0$ an additional
term in the $\chi^2$ sum (\ref{chi}),
\begin{equation}
\left(\frac{0.24-\Omega_m^0}{0.09}\right)^2.\label{cmb}
\end{equation}
This then permits to obtain error contours, and reduces the
correlation coefficients. The value $0.24$ comes from Table 2 of
Tegmark \& al.\,\cite{Tegmark}, who obtained
$\Omega_m^0=0.239^{\,+0.018}_{\,-0.017}$ in a multi-parameter fit to
WMAP and SDSS LRG data. To weaken the effect of this prior we blow
the error up by a factor of 5.

All calculations are done with the classical CERN program MINUIT
(James \& Roos\,\cite{James}) which delivers $\chi^2_{\,best}$,
parameter errors, error contours and parameter correlations. We do
not marginalize, but quote the full, simultaneous confidence region:
a $1\sigma$ error contour in the 3-parametric space then corresponds
to $\chi^2_{\,best} + 3.54$ around the best value $\chi^2_{\,best}$

\begin{figure}[htbp]
\includegraphics{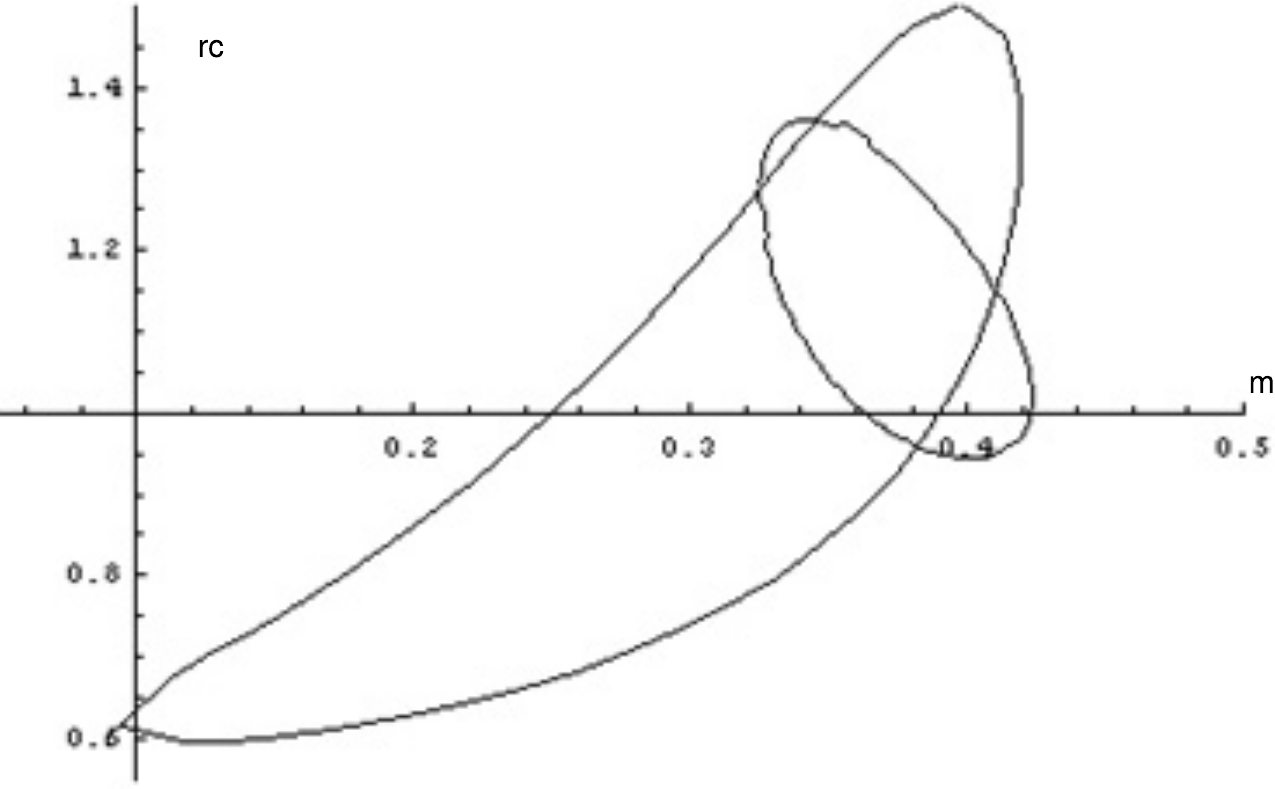}
   \caption{The $1\,\sigma$ confidence region in the
   $(\Omega_m^0,\Omega_{r_c})$-plane from fits to SNeIa data and (i)
   the constraint (\ref{cmb}) (the banana-shaped contour), (ii) the
   constraint (\ref{R})(the elliptic contour)}.
     \end{figure}

With the approximation $x=1$ the best fit parameter values are
\begin{equation}
\Omega_m^0=0.26\pm 0.16,~\Omega_{r_c}=0.82^{+0.69}_{-0.22},~\Omega_A=2.21^{+0.50}_{-0.22}\,,\label{res}
\end{equation}
with $\chi^2=195.5$ for $193-3$ \emph{d.f.} ($\chi^2/d.f.=1.029$),
exactly the goodness-of-fit of the $\Lambda$CDM model.

In Fig.\,1 we plot the corresponding $1\sigma$ confidence region in
the $(\Omega_m^0,\Omega_{r_c})$-plane for a $\chi^2$ sum including
Eqs.\,(\ref{chi}) and (\ref{cmb}), as a banana-shaped contour.
Clearly, the weak prior (\ref{cmb}) has not done much to remove the
degeneracy. The pair of parameters $(\Omega_{r_c},\Omega_A)$ is even
more degenerate (not shown here).

Instead of the rather arbitrary prior (\ref{cmb}), we include a
constraint from the CMB shift parameter $R$, which should not depend
crucially on that it has not been derived in five-dimensional brane
geometry. $R$ is defined by
\begin{equation}
R(\Omega_m,\Omega_{r_c},\Omega_A)=\sqrt{\Omega_m}H_0\int_0^{1089}\frac{dz}{xH(z)}~,
\label{R}
\end{equation}
for which the value $R=1.70\pm 0.03$ has been measured \cite{Wang}.
To permit comparison with the banana-shaped contour, we plot the
SNeIa fit together with the shift parameter (and with x=1), a nearly
elliptically shaped contour, in Fig.\,1.

  \begin{figure}[htbp]
  \includegraphics{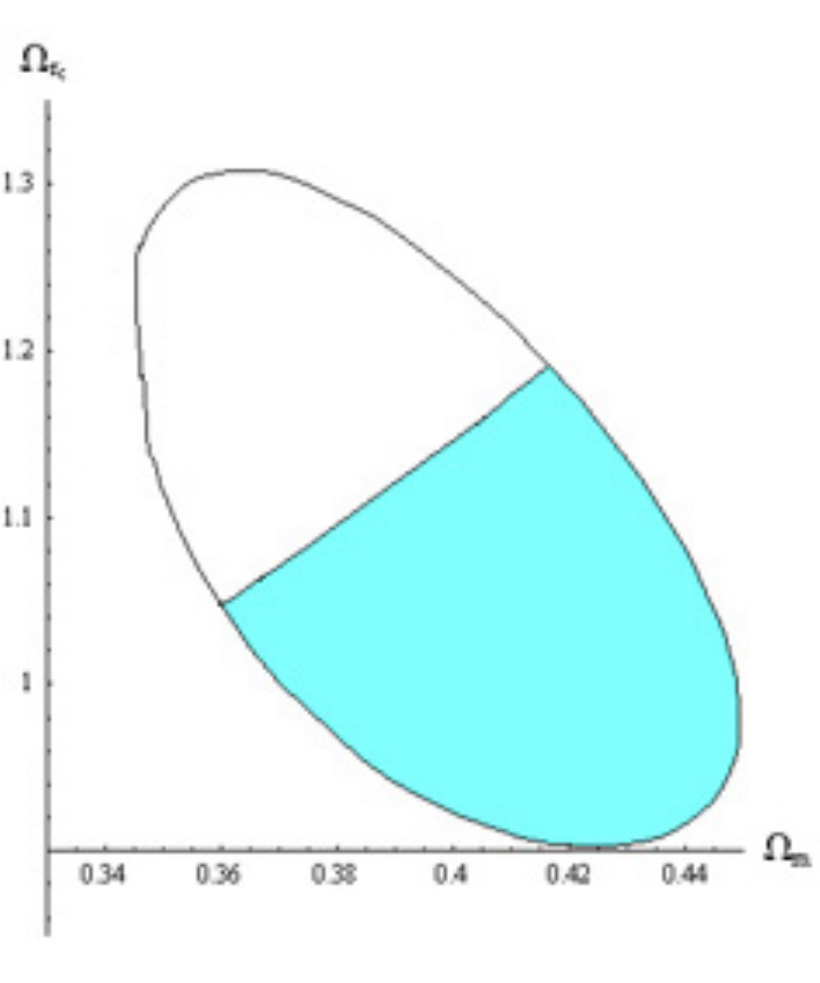}
   \caption{The contour delimits the $1\sigma$ confidence region in
   the $(\Omega_m,\Omega_{r_c})$-plane in a fit to SNeIa data and
   the CMB shift parameter. In the blue region the Universe is
   younger than 12 Gyr.}
  \end{figure}

Fitting next the SNeIa data and $R$ in the same manner as above, but
with the correct $x=0.956$, we find
\begin{equation}
\Omega_m^0=0.40\pm 0.05,~\Omega_{r_c}=1.1\pm 0.2,~\Omega_A=2.5\pm 0.3\,,\label{shift}
\end{equation}
with $\chi^2=195.1$ for $193-3$ \emph{d.f.} ($\chi^2/d.f.=1.027$),
slightly better than the fit (\ref{res}) . We plot in Fig.\,2 the
corresponding confidence region.

The improvement in $\chi^2$ is due to the more exact value of $x$,
mostly felt at small redshifts. The exact position of the ellipse is
dependent on the value of x, as one can see from comparing the
ellipses in Figs.\,1 and 2. The shift in the position of the best
fit from (\ref{res}) to (\ref{shift}), is primarily due to the value
of the new constraint, the shift parameter $R$.

 \begin{figure}[htbp]
 \includegraphics{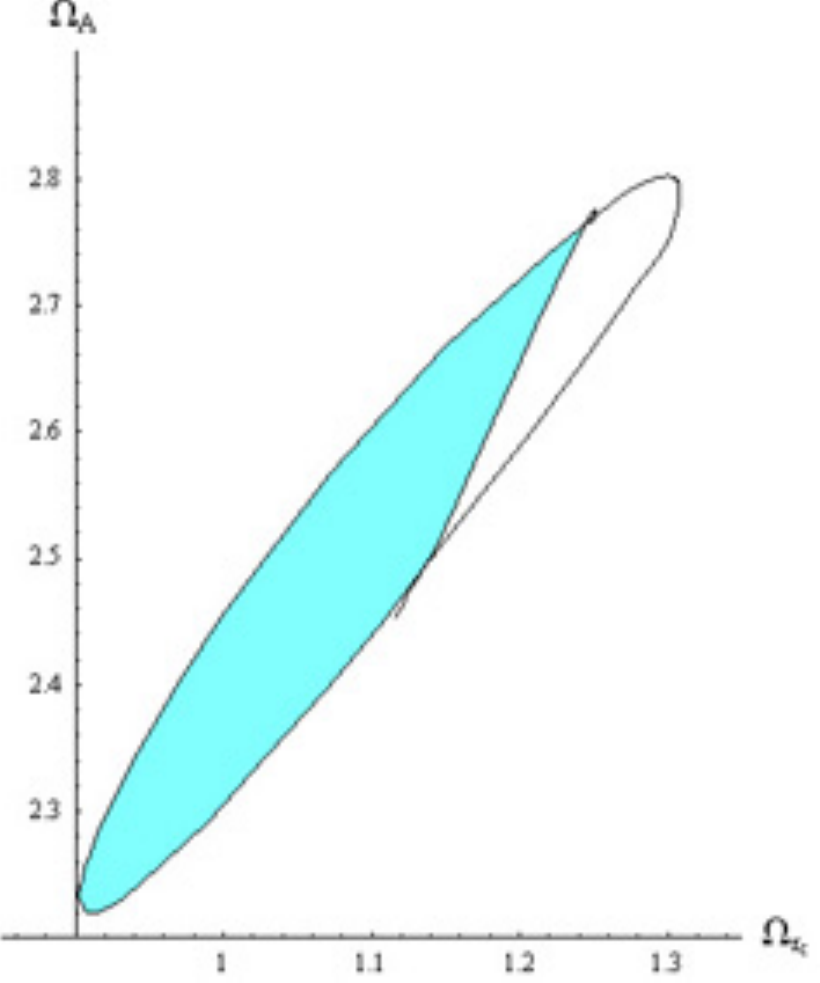}
   \caption{The contour delimits the $1\sigma$ confidence region in
   the $(\Omega_{r_c},\Omega_A)$-plane in a fit to SNeIa data and
   the CMB shift parameter. In the blue region the Universe is
   younger than 12 Gyr.}
  \end{figure}

\section{The age of the oldest star}
In every theory for a universe expanding with velocity $H(z)$, the
age of the universe $t_U$ is given by
\begin{equation}
t_U=\frac{c}{H(1)}\int_0^1\frac{da}{H(a)}.
\end{equation}
The WMAP collaboration \cite{WMAP} quotes $t_U=13.73 \pm 0.12$ Gyr
from a fit of the $\Lambda$CDM model. In the present model $H(a)$ is
given by Eq.\,(\ref{Ha}) with $H(1)=xH_0$.

A model-independent limit of $t_U$ can be obtained from the age of
the oldest star, $t_{\star}$, since $t_U>t_{\star}$, and this limit
can be used to constrain any model for an expanding universe.
Recently, A. Frebel \& al. \cite{Frebel} have reported the discovery
of HE 1523-0901, a strongly r-process-enhanced metal-poor bright
giant star with detected radioactive decay of Th and U. For the
first time, it was possible to employ several different
chronometers, such as the U/Th, U/Ir, Th/Eu, and Th/Os ratios to
measure the age of a star. From 15 such chronometers the weighted
average age of HE 1523-0901 is 13.2 Gyr. Leaving out the Th
chronometers which have the largest systematic errors, the most
useful value is $ t_{\star}=13.4\pm 0.8\,(stat)\pm 1.8\,(syst)\,{\rm
Gyr}.$ Here the systematic error is mainly due uncertainties in the
U production ratio.

The $1\sigma$ statistical error, 0.8 Gyr, can be rewritten as a
one-sided 68\% confidence limit, $t_{\star}>13.0$ Gyr. The
systematic error, 1.8 Gyr, cannot be handled by statistical methods,
so we have to resort to a guess. We opt for constraining our model
by $t_U>12$ Gyr.

The effect of this constraint can be seen in Figs.\,2 and 3. In
Fig.\,2 we over-plot the elliptical contour with the region $t_U<12$
Gyr, where the Universe is too young, painted blue. The value of
$\Omega_A$ along the contour is always the one that minimizes
$\chi^2$ locally.

In Fig.\,3 we plot the $1\sigma$ contour in the
$(\Omega_{r_c},\Omega_A)$-plane with the region $t_U<12$ Gyr painted
blue. The value of $\Omega_m$ along the contour is always the one
that minimizes $\chi^2$ locally. The Universe is too young if the
Chaplygin gas acceleration dominates which it does at large values
of $\Omega_A$, or if the DGP deceleration is too weak which it is at
small values of $\Omega_{r_c}$.

The value of $\Omega_m$ affects $t_U$ in the same way as in the
standard model: the expansion slows down for increasing values, and
the Universe then is younger (blue).

\section{Effective dynamics}
It is of interest to study the effective dynamics of this model, as
expressed by an effective density defined by
\begin{equation}
\rho_{\rm eff}\equiv \rho_{\varphi}-H/\kappa r_c\ ,\label{rhef}
\end{equation}
and an effective equation-of-state parameter
\begin{equation}
w_{\rm eff}\equiv -1-\frac{\dot\rho_{\rm eff}}{3H\rho_{\rm eff}}\ .\label{w}
\end{equation}
Inserting $\rho_{\varphi}$ from Eq.~(\ref{rhoch}) and $H$ from
Eq.~(\ref{Ha}) into Eq.~(\ref{rhef}), one can take time derivatives
to obtain $\dot\rho_{eff}$ in terms of $\dot\rho_{\varphi}$ and
$\dot{H}$. The algebraic expressions for $\rho_{eff}$ and $w_{\rm eff}(z)$ are readily
calculable when parameter values are inserted, but too long to spell out here.

    \begin{figure}[htbp]
\includegraphics{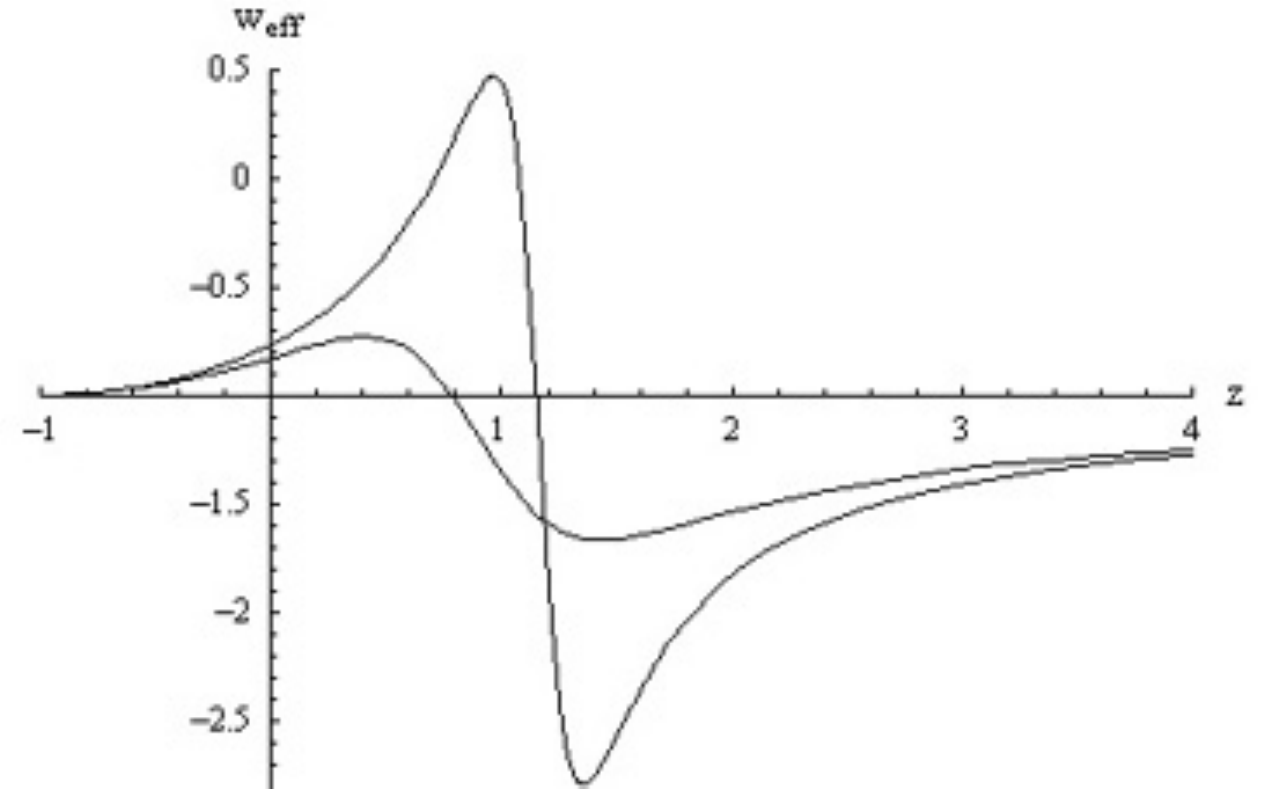}
   \caption{The effective equation-of-state parameter $w_{\rm
   eff}(z)$ as a function of redshift $z$ at two points within the
   $1\sigma$ confidence region.}
    \end{figure}

In Fig.\,4 we show two curves for $w_{\rm eff}(z)$ corresponding to
selected points within the $1\sigma$ confidence range. Both curves
are computed at $\Omega_m=0.344,~ \Omega_A=2.80$; they differ in the
values of $\Omega_{r_c}$: 1.07 and 0.90, respectively. At redshifts
higher than $z\approx 1$, dark energy exhibits phantom-like
acceleration, $w_{\rm eff}<-1$, without phantom matter. In the range
$0<z<1$ (depending on the parameter values) dark energy changes from
phantom-like acceleration across the phantom divide, $w_{\rm
eff}=-1$, to something like quintessence matter, even to $w_{\rm
eff}>0$. At the present time one always has $w_{\rm eff} > -1$
inside the $1\sigma$ confidence range of the parameters, and in the
future $w_{\rm eff}(z)$ always approaches the cosmological constant
value -1.

In most of the parameter space $w_{\rm eff}(z)$ exhibits one or two
mathematical singularities, since $\rho_{\rm eff}$ in
Eq.~(\ref{rhef}) clearly can become temporarily negative. The first
mathematical singularity develops from the peak and dip near $z= 1$
in Fig.\,4, and are located where $\rho_{\rm eff}$ changes sign. The
second singularity, when present, develops near $z=2$.

Another dynamical quantity of interest is the deceleration parameter
$q= -1-\dot H/H^2$. For redshifts $z\approx 4,~q\approx -1.3$, just
as at present. In the range $0<z<4,~q$ goes through a minimum of
$q\approx -1.8$, and in the future it approaches $q=-1$.

\section{Discussion and conclusions}

We have studied a model combining the 2-parametric self-decelerating
DGP model with the likewise 2-parametric standard Chaplygin gas
model. The braneworld DGP model is an example of modified gravity
which is characterized by a length scale $r_cH_0$ which marks the cross-over
between physics occurring in our four-dimensional brane and in a
five-dimensional bulk space. An example of dark energy is Chaplygin
gas which has similar asymptotic properties at early and late cosmic
times, and a characteristic length scale of its own.
We take the two length scales to be proportional, and the
proportionality constant subsequently drops out because of a
normalizing condition at $a=1$. Our model then depends on only one
parameter more than the $\Lambda$CDM model.

The idea to combine the self-decelerating branch of DGP with some
accelerating component has been addressed a few times before in the
literature. Lue \& Starkman \cite{Lue} and Lazkoz \& al.
\cite{Lazkoz} chose the cosmological constant as the accelerating
component, mainly in order to explore models with phantom-like
acceleration, $w<-1$, for large redshifts, but which approach $w=-1$
in the future. The effective acceleration is then increasing with
time as the DGP deceleration vanishes, so that ultimately one
recovers the standard cosmological constant model with all its
conceptual problems. Since the accelerating component is a constant,
it is not characterized by any cross-over scale, nor does this
phantom-like acceleration  ever cross the phantom divide $w=-1$.
Both these papers also discuss observational constraints and
possible future signatures.

Models which at large redshifts exhibit phantom-like acceleration,
and at small redshifts cross the phantom divide, can also be
obtained by replacing the cosmological constant above with a
quintessence field (Chimento \& al. \cite{Chimento}), or as here and
with standard or generalized Chaplygin gas \cite{Bouhmadi} in a
decelerated DGP geometry.

It is easy to explain the coincidence problem in the present model
as well as in the plain DGP model: it is caused merely by the ratio
of the scales of the action, the Planck scale $M_{\rm Pl}$ on our
brane and the bulk scale $M_5$. These constants happen to have
particular time-independent values which determine the DGP
cross-over scale $r_c$.

We find that the effective EOS is phantom-like at large reshifts,
then crosses the phantom divide, so that $w_{\rm eff}(0)>-1$ at the
present time. In the future it approaches $w=-1$.

Our model fits SNeIa data with the same goodness-of-fit as the the
cosmological constant model, it also fits the shift parameter well,
and over a considerable part of the $1\sigma$ confidence range the
age of the Universe is more than 14 Gyr, a constraint derived from
the age of the star HE 1523-0901. In contrast to most other dark
energy models, this model offers a genuine alternative to the
cosmological constant model because it does not reduce to it in any
limit of the parameter space.

Our model should still be tested against other cosmological data,
such as ISW data, CMB and BAO power spectra, all of which has to be
derived in a five-dimensional braneworld cosmology. Such a
derivation has been done recently \cite{ISW}, so that these
constraints can be included in the near future.

\ack It is a pleasure to acknowledge helpful discussions with J.
Nevalainen and J. Ahoranta in Helsinki as well as with R. Durrer and
M. Kunz in Geneva and T. Giannantonio in Portsmouth. T. Davis has
kindly let us use her SNeIa compilation, and H. Ruskeep\"a\"a in
Turku has given precious help with Mathematica.

\end{document}